\documentclass[manuscript]{aastex}

\usepackage{amsmath}
\usepackage{graphicx}
\usepackage{pdflscape}
\usepackage{lineno}
\usepackage[hidelinks]{hyperref}

\slugcomment{\textbf{The Astrophysical Journal Letters}}

\shorttitle{Double Dimorphos}

\begin{document}

\title{A Single Ejection Model of the DART/Dimorphos Debris Trail}

\author{
Yoonyoung Kim$^1$ and David Jewitt$^{1}$
} 
\affil{$^1$Department of Earth, Planetary and Space Sciences, UCLA, Los Angeles, CA 90095-1567}

\email{yoonyoung@epss.ucla.edu}

\begin{abstract}

The collision of the NASA DART spacecraft with asteroid Dimorphos resulted in the formation of a distinctive and long-lived debris trail, formed by the action of solar radiation pressure on ejected particles.  This trail briefly displayed a double appearance, which has been interpreted as the result of a double ejection. We present a model which can  produce a transient double trail without the need to assume a double ejection. Our model explains the appearance of the double trail as a projection of the cone walls when viewed from a large angle to the cone axis and avoids the problem of producing dust in two epochs from a single, instantaneous impact. 
The particles follow a broken power-law size distribution, with differential indices $q=2.7\pm0.2$ (1 $\micron \le a \le$ 2~mm), $3.9\pm0.1$ (2~mm $< a \le$ 1~cm), and $4.2\pm0.2$ (1~cm $< a \le$ 20~cm).
We find that the total trail mass in particles from 1 $\micron$ to 20~cm in size (for an assumed density 3500 kg m$^{-3}$) is $\sim$1.7$\times$10$^{7}$~kg, rising to 2.2$\times$10$^{7}$~kg, when extended to boulders up to 3.5 m in radius.
This corresponds to 0.4--0.6\% of the mass of Dimorphos.

\end{abstract}


\section{Introduction}

The UT 2022 September 26 collision of the NASA DART spacecraft with asteroid Dimorphos resulted in the formation of a long-lived debris trail, formed by the action of solar radiation pressure on solid particles ejected by the impact.   This trail briefly appeared double, but merged into a single trail within two weeks of the impact \citep{Li2023,Opitom2023}. The double trail was interpreted by \cite{Li2023} as evidence for a second discrete ejection event  occurring $\sim$5 to 7 days after the spacecraft impact. \cite{Moreno2023} considered a second ejection $\sim$5.5 days after impact \citep[see also][]{Lin2023}. These authors speculate that a second dust release might be caused by the delayed impact of debris from Dimorphos on the nearby asteroid Didymos, or perhaps by complicated dynamical interactions within the Didymos/Dimorphos binary system. However, it is not obvious why these processes would be lagged from the main impact by nearly a week, given the high speed of ejecta from the impact and the small size of the Didymos/Dimorphos binary system (separation $\sim$1.2 km).

In this Letter, we present an alternate model in which the appearance of the trail is accurately reproduced while the assumption of double ejection is not needed.  Instead, the temporary double trail appearance  is explained as the result of projection from the walls of a conical ejecta curtain, as is commonly observed in laboratory impact experiments \citep[e.g.][]{Melosh1989}.  This projection effect fades with the rapidly changing observing geometry (including a three-fold increase in the geocentric distance and a corresponding loss of spatial resolution) and as the trail progressively spreads, leading to a single trail morphology as observed at later times.  We use our model in conjunction with Hubble Space Telescope (HST) data (Table \ref{geometry}) to estimate the particle size distribution and the total mass of material released by the DART/Dimorphos impact.

\section{Model Description}

The  Dimorphos debris cloud  exhibited three main structures: (1) an impact ejecta cone consisting of short-lived and irregular structures, (2) a diffuse envelope, and (3) a radiation pressure swept debris trail with a temporarily double appearance \citep{Li2023}.  In addition, dozens of impact-produced boulders up to $\sim$7 m in diameter were detected in deep HST images \citep{Jewitt2023}.  Our primary objective is to model the radiation pressure swept trail.

To model the morphology of the  debris trail, we used a Monte-Carlo simulation of the dust \citep{Ishiguro2007,Kim2017}. 
Previous applications of this model showed that projection from the walls of an impact cone can produce a double-tail morphology \citep{Ishiguro2011,Kim2017b}. Given the morphological similarity with the DART/Dimorphos trail, we consider a similar model. Specifically,
we assume that the dust particles were impulsively launched in a conical ejecta curtain \citep[c.f.][]{Melosh1989},  on UT 2022 September 26, 23:14:24.183 \citep{Daly2023}.
We follow the motions of ejected particles in the curtain under the action of solar gravity and radiation pressure, following a procedure developed in \cite{Ishiguro2011}. We consider the conical curtain to be symmetric with respect to a vector perpendicular to the asteroid surface ($\alpha_\mathrm{cone}$, $\delta_\mathrm{cone}$) with a half-opening angle of $\theta$ (i.e.~the vertex angle of the cone is 2$\theta$). The angular thickness of the cone walls is written $\delta \theta$.  Particles are ejected uniformly in the angle range $\theta\pm(\delta\theta/2)$, with no particles outside this range.

The number of dust particles within a radius range from $a$ to $a$+$da$ is given by a power-law  distribution with size index $q$, $N(a)da = N a^{-q}~ da$, where $N$ is the reference number of dust particles.  We further represent the ejection terminal velocity as a power law function of the  particle radius,  such that $v_{\infty} = V_0 a^{-k} v$. $V_0$ is the reference ejection velocity (m s$^{-1}$) of  particles having $a$=1~m, and $k$ is the velocity power index. The dimensionless random variable $v \ge 0$ follows a Gaussian probability density function given by $P(v) = 1/(\sqrt{2\pi}\sigma_v)\exp (- (v-1)^2/{2\sigma_v^2})$, where we set standard deviation $\sigma_v$=0.3 \citep{Ishiguro2013}.

\begin{figure}
\epsscale{0.95}
\plotone{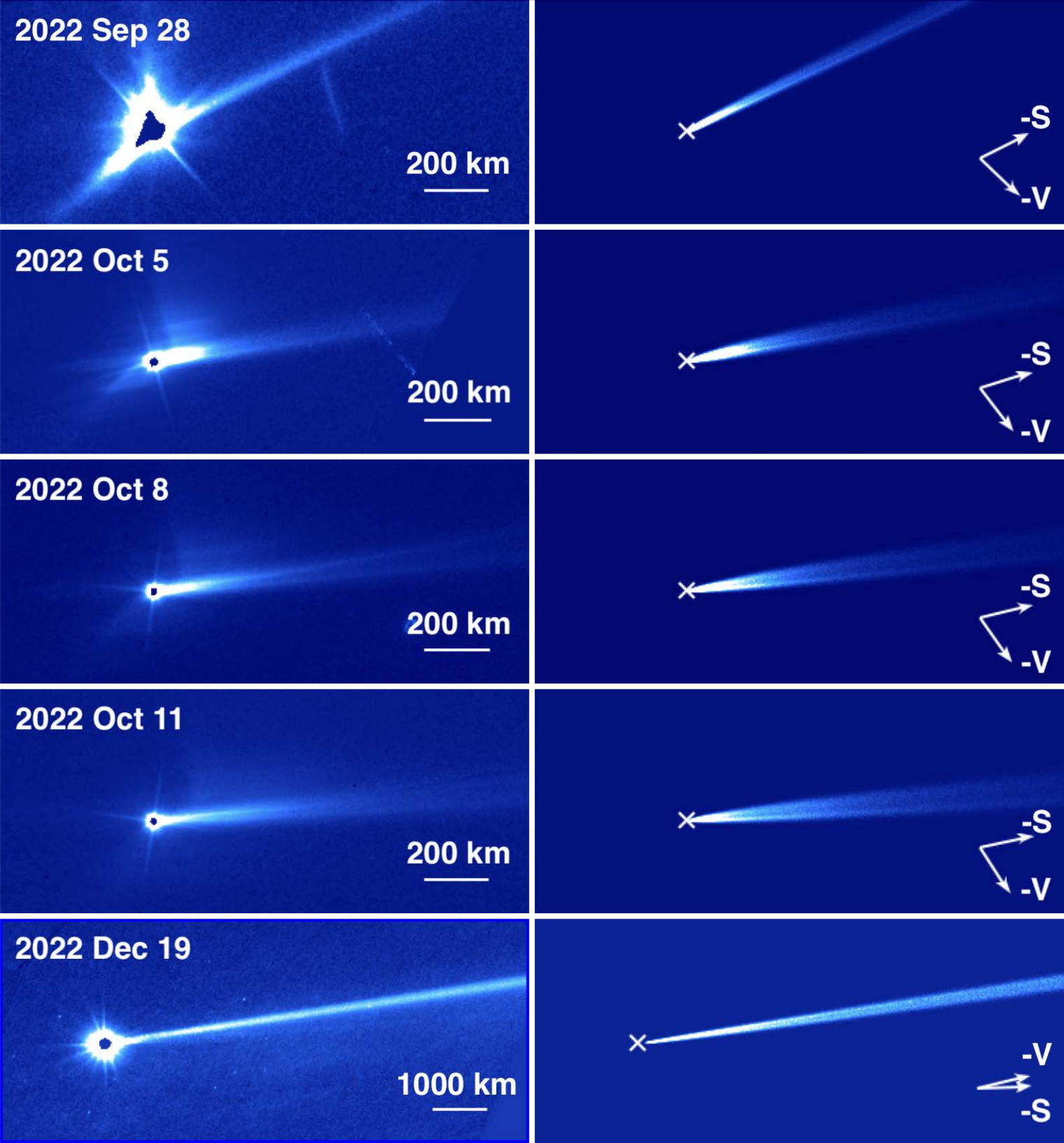}
\caption{Comparison between HST images (left) and Monte Carlo models (right). A linear scale bar, the projected antisolar direction ($-S$), and the negative heliocentric velocity vector ($-V$) are indicated. The Dimorphos location is marked with a cross in each panel.  Note the scale change in the bottom two panels.
 \label{model}}
\end{figure}

The trajectories of the particles were computed from the terminal velocity and the ratio of radiation pressure acceleration to solar gravity, $\beta$.
For spherical particles, $\beta$  is approximately given by $\beta = 570/(\rho a_{\micron})$, where $\rho$ is the mass density of dust particles in kg m$^{-3}$ and $a_{\micron}$ is in microns \citep{Burns1979}.
The density of the ejecta particles is not known but, for ease of comparison with the work of   \cite{Moreno2023}, we assumed $\rho$ = 3500 kg m$^{-3}$. 
Because of the very small observer distance (0.076 au at the moment of impact), a small change in the observer position could have a significant effect on the morphology, so we used HST's precise position at a given time obtained from JPL Horizons. We performed multiple simulations with different parameter sets and then visually compared the resulting model images to the data to identify plausible parameters. The best-fit parameters are given in Table \ref{parameter}. Figure \ref{model} compares the observations with the best-fit models.

\section{Discussion}

\subsection{Morphology}

The parameter ranges to be explored were selected based on previous studies of active comets and asteroid dust trails \citep{Ishiguro2007,Kim2017b,Agarwal2023}.  \cite{Moreno2023} employed a piece-wise model of Dimorphos in which the dust particles were ejected in two different velocity ranges (fast and slow), the latter in two epochs in order to generate two trails.  Fast particles have $V \gg V_e$, where $V_e \sim$0.24 m s$^{-1}$ is the system escape speed, while slow particles have $V \lesssim V_e$.  They also employed a power law representation of the particle size distribution, and required two different power laws with a break at $a \sim$ 3 mm in order to fit the data.  Here, we ignore the high velocity components associated with the earliest observations of the impact debris and model only the trail particles, which are ejected slowly enough that radiation pressure dominates their motion. Specifically, we ignore particles with ejection speeds $V \gg V_e$ because these particles do not contribute to the trail we seek to model.

We first used the cone axis solutions (141\arcdeg, 25\arcdeg) and (120\arcdeg, 10\arcdeg) from \cite{Li2023} to find the ejection terminal velocity and cone angle parameter sets, fitting the double trail position angle and trail width of the HST data. We assume that the cone is symmetric about its long axis; the real cone may include asymmetries that alter the appearance slightly when projected into the plane of the sky.  The double trail appearance was easily obtained assuming low ejection velocities.
We derived $V_0 \sim$ 1.5 mm s$^{-1}$ (for $a$ = 1~m particles) and $k \sim 0.5$, indicating that the maximum speed was $\sim$1.5 m s$^{-1}$ for $a$ = 1~\micron~particles. At $k<0.5$, the wedge angle of the October 8 data could not be matched, and at $k>0.5$, the trail in the December 19 data widened too steeply.  High velocity impact experiments show a weak and inverse relation between fragment speed and size, consistent with the small exponent derived here \citep{Nakamura1991,Giblin1998}.

The smallest particle size (1~\micron) was determined from the length of the trail in the earlier images. The largest particle size ($a \gtrsim$ 0.2~m) was obtained from the absence of a clear gap between the nucleus and the trail in the latest images.
Terminal velocity $v_{\infty}$ is the excess speed after particles climbed out of the potential well of the system, and the actual ejection speed $U$ is given by $U = \sqrt{v_{\infty}^2+V_e^2}$, where $V_e$ is the gravitational escape speed from the system. Over the size range 1 $\micron < a < 0.2$~m, the ejection speed $U$ varies from 0.24 to 1.5 m s$^{-1}$.

\begin{figure}
\epsscale{0.9}
\plotone{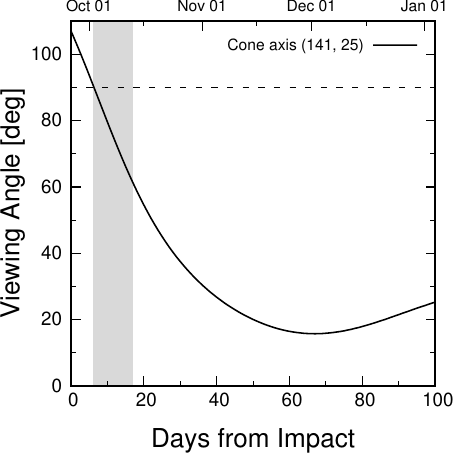}
\caption{Viewing angle (angle between the trail axis and the line of sight) as a function of time, expressed as days after impact.
The shaded region indicates when the double tail was observed.
The ejecta cone is viewed edge-on when viewing angle = 90\degr~(dashed line), at which point the walls of the hollow cone are most clearly visible.
 \label{view}}
\end{figure}

Unless the cone axis deviates significantly from the solutions used in \cite{Li2023} and \cite{Cheng2023}, no significant difference appears in the morphology. We find a best-fit half-opening angle of $\theta=60\degr\pm10\arcdeg$ (in excellent agreement with $\theta=62\degr\pm5\arcdeg$ from Li et al.~2023) with the cone axis at $120\arcdeg \lesssim \alpha_\mathrm{cone} \lesssim 145\arcdeg$ and $10\arcdeg \lesssim \delta_\mathrm{cone} \lesssim 25\arcdeg$.    Assuming the cone axis at (141\arcdeg, 25\arcdeg), Figure \ref{view} shows the angle between the trail axis and the line of sight, measured from the nucleus. Interestingly, the viewing angle reached 90\degr~in early October when the double tail was first reported \citep{Li2023,Murphy2023}, and maintained a large angle during the period when the double tail was observed. The double trail was not noticed in the first week after impact presumably because the ejecta cone had not expanded enough for the walls to appear separated (c.f.~Figure \ref{model}). 

As such, our model explains the temporary appearance of the double trail as a projection of the cone walls when viewed from a large angle to the cone axis. 
The double appearance faded after mid-October as the observing geometry changed rapidly (towards a larger observer distance, poorer spatial resolution and smaller cone viewing angle) and the particles dispersed by radiation pressure progressively filled the cone. The latter takes about 70 days to fill the cone, after which a single trail morphology appears regardless of the viewing geometry.

\subsection{Size Distribution and Mass}

\begin{figure}
\epsscale{1.0}
\plotone{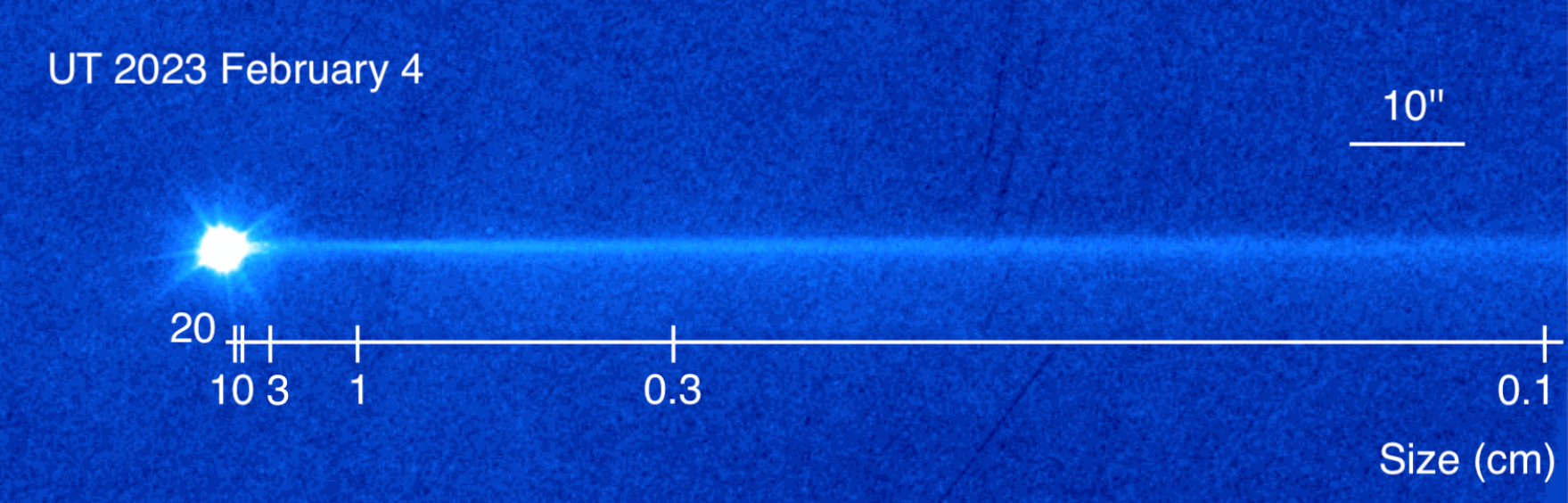}
\caption{Composite HST image from UT 2023 February 4, rotated so that the trail lies horizontally. We mark the distances reached by dust particles under the action of radiation pressure, as a function of their radius (in centimeters).
 \label{amax}}
\end{figure}

The particle size dependence of the radiation pressure parameter, $\beta$, imposes a size gradient along the length of the trail, from large particles near the impact site to small ones far away.  Combined with measurements of the trail surface brightness as a function of position, this size gradient can be used to determine the particle size distribution in the ejected material.  For reference, Figure~\ref{amax} shows an HST image taken on UT 2023 February~4 annotated to show the approximate  distances along the trail to which  particles of a given size have been swept by radiation pressure in the $\sim$4 months since impact.  Evidently, particles smaller than $a \sim$1~mm have already been removed from the HST field of view by this date under the action of solar radiation pressure.

We used HST observations (Table \ref{geometry}), in order to determine the surface brightness of the trail. The nucleus and surroundings were saturated in most early HST images and, consequently, we did not measure or fit within a few arcseconds of the nucleus. As noted by \cite{Li2023} and \cite{Moreno2023}, no single power law can fit the surface brightness measurements.  In agreement with this conclusion, our models yield  $q = 2.7\pm0.2$ for 1 $\mu$m $\le a \le$ 2 mm and $q = 3.9\pm0.1$ for 2 mm $< a \le$ 1 cm.  

\begin{figure}
\centering
\epsscale{1.0}
\plotone{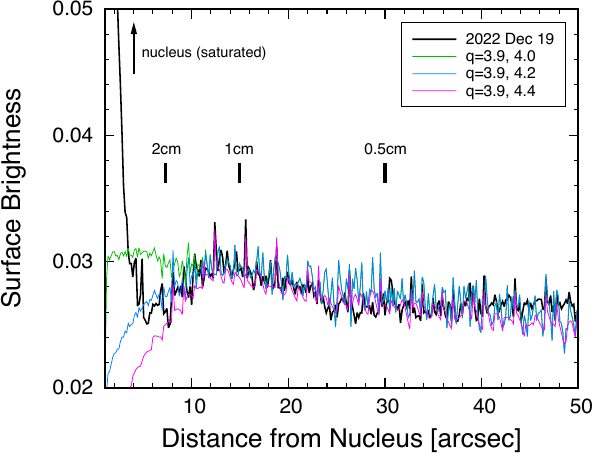}
\caption{The surface brightness vs.~distance along the trail (black line; 0.001 = 27.2 mag.~arcsec$^{-2}$).
Lines show models for a  power law, broken at $a$=1~cm, with differential indices $q$=3.9 and 4.0 (green), $q$=3.9 and 4.2 (blue), and $q$=3.9 and 4.4 (magenta). Approximate locations of particles as a function of their radius are marked.  The nucleus was saturated in the data and was not fitted.  
 \label{sbr}}
\end{figure}

In addition, the surface brightness profile close to the nucleus allows us to place a constraint on   particles even larger than 1 cm. We find that models with $q < 4.0$ leave too many large particles within $\sim$5 to 12\arcsec~of the nucleus, where the data show  a surface brightness downturn (Figure \ref{sbr}).  Conversely, power laws with $q \gtrsim 4.4$ are inconsistent with the data by under-predicting the near-nucleus surface brightness, leaving $q = 4.2\pm0.2$ as our best estimate of the large particle index.  We conclude that particles larger than $\sim$1 cm are distributed according to a steeper power law than smaller particles.  

In summary, the best-fit size distribution obtained from our models is

\begin{equation}
N(a)da =
\left \{
\begin{array}{lcl}
N_1~ a^{-(2.7\pm0.2)}~ da & \hspace{5mm} \mathrm{for} & 1\times10^{-6}\leq a \leq 2\times10^{-3} \\
N_2~ a^{-(3.9\pm0.1)}~ da & \hspace{5mm} \mathrm{for} & 2\times10^{-3} < a \leq 1\times10^{-2} \\
N_3~ a^{-(4.2\pm0.2)}~ da & \hspace{5mm} \mathrm{for} & 1\times10^{-2} < a \leq 2\times10^{-1}
\end{array}
\right .
\label{na}
\end{equation}

\noindent where $a$ is expressed in meters, and $N_1$, $N_2$, and $N_3$ represent the reference dust production rates.  

For comparison, \cite{Li2023}  found a break in the size distribution for particle radii of a few millimeters, with $q$ = 2.7$\pm$0.2 for smaller dust and $q$ = 3.7$\pm$0.2 for larger particles.   \cite{Moreno2023} fitted a broken power-law with $q=2.5$ for particles $a\lesssim3$~mm, and with a higher slope of $q=3.7$ for particles $a\gtrsim3$~mm.  These results are concordant with those in Equation \ref{na}.  Broken power law size distributions have also been reported in the debris from natural asteroid breakups \citep{Ye2019,Jewitt2019,Jewitt2021}.  

The important feature of Equation \ref{na} is that it shows that the size distribution steepens towards larger particle sizes.  
We note that the choice of the largest particle size ($a_\mathrm{max}$) does not have a significant effect on the near-nucleus surface brightness, indicating that $a_\mathrm{max}$ could be much larger than 0.2~m, even extending to the size of the boulders ($a_\mathrm{max} \lesssim$ 3.5 m; Jewitt et al.~2023).

For a collection of spheres, the total cross section, $C_d$, is given by

\begin{equation}
C_d = \int_{a_\mathrm{min}}^{a_\mathrm{max}} \pi a^{2} N(a) da~,
\end{equation}

\noindent where we derive the reference number $N_1$, $N_2$, and $N_3$ from the early photometric result by substituting $N(a)da$ in Equation \ref{na}. The total mass can be estimated from the early photometry (i.e. brightening) combined with the carefully determined size distribution.
\cite{Graykowski2023} reported brightening on impact ($\Delta C \sim$ 3.82 km$^2$ increase in cross-sectional area from the peak to the nucleus level, and $\Delta C \sim$ 1.64 km$^2$ after the fast-moving ejecta moved out of the photometric aperture).
Substituting $\Delta C$ = 1.64 km$^2$ gives $N_2$=480, where $N_1=N_2 \times 0.002^{(2.7-3.9)}$ and $N_3=N_2 \times 0.01^{(4.2-3.9)}$.

The total mass of the dust cloud, $M_d$, is given by

\begin{equation}
M_\mathrm{d} = \int_{a_\mathrm{min}}^{a_\mathrm{max}} \frac{4}{3}\pi \rho_d a^3 N(a)da~,
\end{equation}

\noindent where we find that the total mass of the trail in the particle size range from 1 $\micron$ to 0.2 m in size, is $\sim$1.7$\times$10$^{7}$ kg (assumed density 3500 kg m$^{-3}$), corresponding to about 0.4\% of the mass of Dimorphos.
Extending this size distribution to the largest observed boulder radius 3.5 m \citep{Jewitt2023}, the total escaping ejecta mass from the DART spacecraft impact is estimated to be at least 2.2$\times$10$^{7}$ kg, equal to about 0.6\% of the mass of Dimorphos.
This value is compared to the total boulder mass $M_b=8\times$10$^{6}$ kg (scaled to $\rho$ = 3500 kg m$^{-3}$, \cite{Jewitt2023}).  We note that the observed boulders reported in \cite{Jewitt2023} are not part of the radiation pressure dominated trail, and there may be uncertainty in the actual boulder size and fallback mass.

We calculate the fallback mass of large dust particles within a size range of $a_\mathrm{max}$ (largest trail particle observed) and $a_\mathrm{b}$ (largest boulder observed),

\begin{equation}
M_\mathrm{fb} = \int_{a_\mathrm{max}}^{a_\mathrm{b}} \frac{4}{3}\pi \rho_d a^3 N(a)da~,
\end{equation}

\noindent where we set $a_\mathrm{max}$ = 0.2~m and $a_\mathrm{b}$ = 3.5~m to find $M_\mathrm{fb} \sim 5.4\times10^{6}$ kg. This fallback material is enough to coat the entire surface of Dimorphos to a depth of $D = M_\mathrm{fb}/(4\pi r_n^2 \rho)\sim$2 cm, and could be contained within a hemispherical cavity $\sim$15 m in radius. All of the above mass estimates are subject to potentially considerable (factor of two or larger) systematic uncertainties, since they rely upon unmeasured particle densities, albedos and phase function.

One difference between the present study and those of \cite{Li2023} and \cite{Moreno2023} is our use of data from a larger range of dates.  We model HST imaging  obtained up to 2023 April 10 (i.e.~7 months from impact) whereas \cite{Li2023} used HST data taken up to 2022 October 14 (3 weeks from impact) and \cite{Moreno2023} modeled the same HST data  supplemented by ground-based images extending to 2022 December 24 (i.e.~3 months from impact). The later observations (Table \ref{geometry}) allow more time for the largest, slowest particles to move away from Dimorphos. They provide the best evidence for  the lack of a clear gap near the impact site and so for the existence of a second break in the size distribution. 

A second difference is that our model does not require the assumption of a delayed ejection of unclear origin, occuring nearly a week after the DART impact.  Photometry places an independent constraint on secondary dust ejection because a separate  mass-loss event should cause a delayed trail brightening in proportion to the added cross-section.  For example, \cite{Moreno2023} infer that the masses of the primary and secondary trails are in the ratio 4.3:3.0 (c.f.~their Table 4).  With similar size distributions in the two trails, this would also be the ratio of the cross-sections.  The emergence of the secondary trail should then increase the total trail brightness by 3/4.3 $\sim$70\% starting 5 to 7 days after the impact.

\begin{figure}
\epsscale{0.85}
\plotone{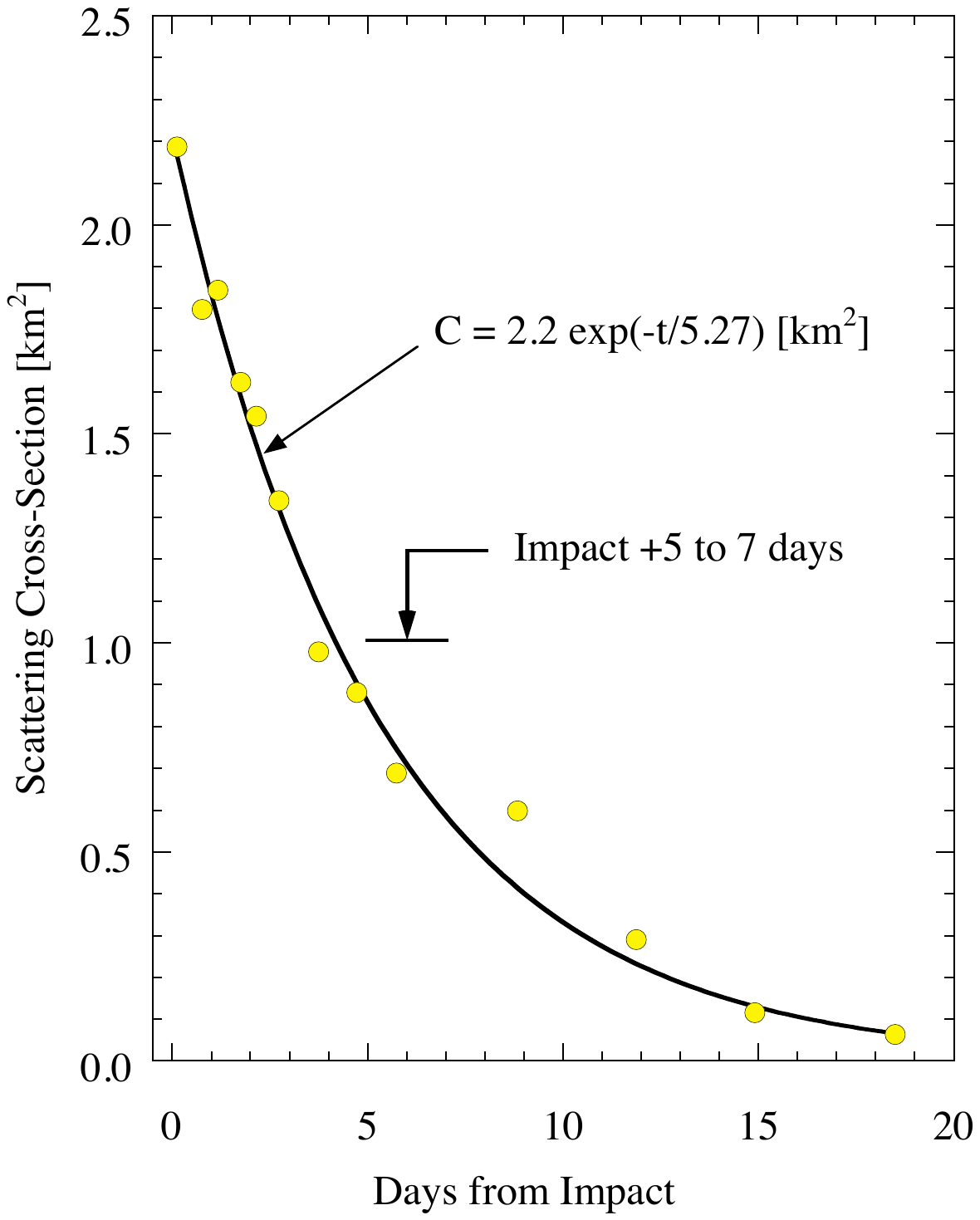}
\caption{Scattering cross-section of the debris trail within a 50 km radius projected aperture, as a function of time.  The solid black line shows an exponential fit to the data. The secondary tail ejection time from the model of \cite{Li2023} is indicated. 
 \label{photometry}}
\end{figure}

Figure \ref{photometry} shows the scattering cross-section of the debris trail within 50 km of Dimorphos. We used the aperture photometry from Extended Data Figure 4B of \cite{Li2023}, from which the contribution to the photometry from Didymos/Dimorphos has been subtracted.  We assumed a geometric albedo 0.1 
(for easy comparison with \cite{Moreno2023})
\footnote{\cite{Naidu2020} measured albedo 0.15$\pm$0.04.  We examined NEOWISE data in search of  thermal emission from the system, finding images at 3.4 $\micron$ and 4.6 $\micron$. However, the NEOWISE signal-to-noise ratio is insufficient to accurately estimate the albedo.}
and a phase function of 0.04 magnitudes degree$^{-1}$ in order to calculate the cross-section from the photometry.  We also show an exponential function, $C = C_0\exp(-t/T)$, least-squares fitted to the data. The fit matches the data well given initial cross-section $C_0 = 2.2$ km$^2$ and e-folding decay time $T = 5.27\pm0.01$ days.  The cross-section  shows no evidence for an increase at the nominal time of the second ejection (impact +5 to 7 days, marked in the Figure by a short horizontal bar). Instead, the cross-section is larger than the fitted value only at $t$ = 8.8 days but returns to the exponential decay by $t$ = 11.8 days.  Independent ground-based data provide no support for a brightening at 8.8 days \citep{Graykowski2023,Moreno2023} but are subject to larger uncertainties than the HST measurements. We conclude that published evidence for a post-impact trail brightening is limited to a single HST measurement from $t$ = 8.8 days and that, while the available photometric data do not absolutely rule out the possibility of a delayed brightening of the trail, neither do they  provide  compelling evidence for it. 
 More and better photometric data are required.    
We thus believe that our single ejection model is consistent with the available data.

We end with a reminder that Monte Carlo models are necessarily non-unique. In this sense, the broad agreement between the particle size distributions presented here and those by Li et al.~and Moreno et al.~is encouraging. The distinction of the current model is that it satisfies Occam's Razor, by not requiring the assumption of a second ejection of uncertain origin in order to produce the double trail.

\clearpage
\section{Summary}

We present a single ejection model for the DART-produced double debris trail of asteroid Dimorphos, with the following results:

\begin{enumerate}

\item The combined HST dataset can be matched by a size distribution with three segments, with differential power-law indices $q_1=2.7\pm0.2$ (1 $\micron \le a \le$ 2~mm), $q_2=3.9\pm0.1$ (2~mm $< a \le$ 1~cm), and $q_3=4.2\pm0.2$ (1~cm $< a \le$ 20~cm).

\item The ejected mass in particles up to 3.5 m in radius (i.e.~the size of the largest ejected boulder) is 2.2$\times$10$^{7}$ kg (assumed density 3500 kg m$^{-3}$), corresponding to about 0.6\% of the mass of Dimorphos.  Systematic uncertainties on the ejected particle mass are likely a factor of two, or more, depending especially on the unmeasured values of the particle density, albedo and phase function.

\item Our model explains the temporarily double appearance of the  trail as a projection of the ejection cone walls when viewed from a large angle to the cone axis.  A delayed ejection is not needed to explain the second trail.

\end{enumerate}

\acknowledgments
We thank the anonymous referee for a prompt review and Jian-Yang Li for comments. Based on observations made with the NASA/ESA Hubble Space Telescope, obtained from the data archive at the Space Telescope Science Institute. STScI is operated by the Association of Universities for Research in Astronomy, Inc. under NASA contract NAS 5-26555. Support for this work was provided by NASA through grant numbers GO-17289, GO-17293, and GO-17297 from the Space Telescope Science Institute, which is operated by auRA, Inc., under NASA contract NAS 5-26555. 

{\it Facilities:}  \facility{HST}.

\clearpage

\begin{deluxetable}{lcrrrrcc}
\tabletypesize{\scriptsize}
\tablecaption{Observations Used
\label{geometry}}
\tablewidth{0pt}
\tablehead{\colhead{UT Date and Time}   &  \colhead{$\Delta T_i$\tablenotemark{a}} & \colhead{$r_H$\tablenotemark{b}}  & \colhead{$\Delta$\tablenotemark{c}} & \colhead{$\alpha$\tablenotemark{d}}  & \colhead{$\theta_{- \odot}$\tablenotemark{e}} & \colhead{$\theta_{-V}$\tablenotemark{f}}  & \colhead{$\delta_{\oplus}$\tablenotemark{g}}  }

\startdata
2022 Sep 26 23:14 (Impact)   &  0.00 & 1.046 & 0.076 & 53.2 & 297.9 & 228.1 & 47.6 \\
2022 Sep 27 07:25--07:44 & 0.35 & 1.045 & 0.075 & 53.7 & 297.3 & 227.7 & 47.7 \\
2022 Sep 27 16:57--17:31 & 0.75 & 1.044 & 0.075 & 54.4 & 296.7 & 227.2 & 48.0 \\
2022 Sep 28 02:28--03:02 & 1.14 & 1.043 & 0.074 & 54.9 & 296.0 & 226.6 & 48.1 \\
2022 Oct 2 16:01--16:35 & 5.71 & 1.033 & 0.071 & 61.7 & 289.0 & 220.9 & 48.4 \\
2022 Oct 5 18:38--19:12 & 8.82 & 1.027 & 0.071 & 66.0 & 285.4 & 217.6 & 46.7 \\
2022 Oct 8 19:40--20:15 & 11.86 & 1.022 & 0.073 & 69.6 & 282.9 & 215.2 & 43.8 \\
2022 Oct 11 20:42--21:16 & 14.90 & 1.018 & 0.075 & 72.5 & 281.5 & 213.6 & 40.1 \\
2022 Dec 19 15:05--20:25  & 83.78 & 1.177 & 0.219 & 25.5 & 271.5 & 282.9 & -3.8 \\
2023 Feb 4 13:30--Feb 5 18:38  &  131.20 & 1.433 & 0.496 & 21.0 & 110.1 & 274.2 & -6.1 \\
2023 Apr 10 10:52--Apr 11 22:12  & 196.22 & 1.771 & 1.282 & 33.7 & 104.2 & 280.6 & -2.1 \\
\enddata

\tablenotetext{a}{Number of days from impact}
\tablenotetext{b}{Heliocentric distance, in au}
\tablenotetext{c}{Geocentric distance, in au}
\tablenotetext{d}{Phase angle, in degrees }
\tablenotetext{e}{Position angle of projected anti-solar direction, in degrees }
\tablenotetext{f}{Position angle of negative heliocentric velocity vector, in degrees }
\tablenotetext{g}{Angle from orbital plane, in degrees}

\end{deluxetable}

\clearpage

\begin{deluxetable}{lll}
\tablecaption{Input and best-fit parameters for the dust ejection model. \label{parameter}}
\tablewidth{0pt}
\tablehead{Parameter & Input values & Best-fit values}
\startdata
$\theta$ [\arcdeg]        & 40 -- 90 with 5 interval & 60$\pm$10 \\
$\delta \theta$ [\arcdeg]  & 10 -- 40 with 5 interval & 30$\pm$10 \\
$\alpha_\mathrm{cone}$ [\arcdeg] & 0 -- 360 with 5 interval & 120--145 \\
$\delta_\mathrm{cone}$ [\arcdeg] & -90 to 90 with 5 interval & 10--25 \\
$V_0$ [m/s] & 0.001-- 0.240 with 0.0005 interval & 0.0015$\pm$0.0005 \\
$k$ & 0.1 -- 0.6 with 0.1 interval & 0.5$\pm$0.1 \\
$q_1$ & 2.5 -- 4.5 with 0.1 interval & 2.7$\pm$0.2 (1 $\micron \le a \le$ 2~mm) \\
$q_2$ & 2.5 -- 4.5 with 0.1 interval & 3.9$\pm$0.1 (2~mm $< a \le$ 1~cm) \\
$q_3$ & 2.5 -- 4.5 with 0.1 interval & 4.2$\pm$0.2 (1~cm $< a \le$ 20~cm) \\
\enddata
\end{deluxetable}

\clearpage

\clearpage

\clearpage

\clearpage

\clearpage

\end{document}